\documentstyle{article}
\begin{document}
\title{Off-shell indefinite-metric triple-bracket\\
generalization of the Dirac equation
\footnote{To be published in proceedings of {\it 
XXI Colloqium on Group Theoretical Methods in 
Physics\/}, July 15--20, 1996, Goslar, Germany}}
%
\author{Marek Czachor and Maciej Kuna\\
Wydzia{\l}  Fizyki Technicznej i Matematyki Stosowanej\\
 Politechnika Gda\'{n}ska,
ul. Narutowicza 11/12, 80-952 Gda\'{n}sk, Poland}
%
\newcommand{\pp}[1]{\phantom{#1}}
\newcommand{\be}{\begin{eqnarray}}
\newcommand{\ee}{\end{eqnarray}}
\newcommand{\ve}{\varepsilon}
\newcommand{\Tr}{{\rm Tr\,}}
\newtheorem{th}{Theorem}
\newtheorem{lem}[th]{Lemma}


\maketitle
\begin{abstract}
We present an off-shell indefinite-metric reformulation of the
earlier on-shell positive-metric triple bracket generalization
of the Dirac equation \cite{c1,c2}. The new version of the
formalism  solves the question of its manifest covariance.
\end{abstract}

\section{Hamilton, Lie-Poisson and Lie-Nambu versions
of the off-shell Dirac equation}

In linear and pure-state
case the standard positive metric
associated with the Dirac equation is constructed by means of a 
spacelike hyperplane $\Sigma$ but the continuity equation guarantees that
the metric is in fact $\Sigma$-independent. In the generalized
density-matrix nonlinear formulation we cannot use this argument
and hence the independence of the whole formalism from the
choice of $\Sigma$ is unclear. The natural way out of the
difficulty is to simply use the indefinite metric formulation
which does not depend on any hyperplane. We therefore obtain 
a formalism which is manifestly covariant. The convention we use
assumes that repeated Greek indices imply simultaneous
summation over bispinor and integration over spacetime indices.

We begin with the off-shell version of the spinor form of the 
(free) Dirac equation
\be
{\sqrt{2}}\nabla{_{B}}{_{A'}}\psi^B= \partial_{s}\psi_{A'};
\quad
{\sqrt{2}}\nabla{_{A}}{_{B'}}\psi^{B'}=-\partial_{s} \psi_{A}.
\ee
Here $\partial_{s}$ denotes a partial derivative with respect to
a ``proper
time" which is conjugate to mass \cite{arg1}. 
The Hamiltonian function (``average mass") is given by
\be
H=
{\sqrt{2}}
\int d^4x
\Bigl(
\bar \psi^{A'}i\nabla{_{B}}{_{A'}}\psi^B
+
\bar \psi^{A}i\nabla{_{A}}{_{B'}}\psi^{B'}
\Bigr)
=
\int d^4x 
\bar \psi^{\alpha'}(x)g{_{\alpha'}}{^{\beta}}
i\nabla{_{\beta}}{^{\gamma}}\psi_{\gamma}(x)
=\bar H
\ee
and leads to Hamilton equations equivalent to the Dirac
equation:
\be
i\partial_{s}
\psi_{\alpha}
=
-
g{_{\alpha}}{_{\alpha'}}
\frac{\delta H}{\delta \bar \psi_{\alpha'}};\quad
i\partial_{s}
\bar \psi_{\alpha'}
=
g{_{\alpha}}{_{\alpha'}}
\frac{\delta H}{\delta \psi_{\alpha}}.
\ee
The abstract index bispinor convention is explained in the Appendix.
The Poisson tensor  and the symplectic
form are given by $I_a=-g_{\alpha}{_{\alpha'}}$ and
$\omega^a=-g^{\alpha}{^{\alpha'}}$ respectively. 
With these identifications
and following step by step the scheme discussed in \cite{c1,c2} 
we obtain the Lie-Poisson and Lie-Nambu structures in their
off-shell and indefinite-metric form. 

\subsection{Metric and higher order metric tensors}

Metric tensors allowing to raise and lower indices in
the infinite-dimensional Lie algebra are 
\be
g^{ab}=
g{^{\alpha}}{^{\beta'}}g{^{\beta}}{^{\alpha'}}
\delta(a-b')
\delta(a'-b);
\quad
g_{ab}=
g{_{\alpha}}{_{\beta'}}g{_{\beta}}{_{\alpha'}}
\delta(a-b')
\delta(a'-b).
\ee
The two tensors are symmetric and satisfy
$
g^{ab}g_{bc}=
\ve{_{\gamma}}{^{\alpha}}
\ve{_{\gamma'}}{^{\alpha'}}
=\ve_c{^a}.
$
Skipping the Dirac deltas we define higher order metric tensors 
which will be used in Casimir invariants:
\be
g^{a_1\dots a_n}&=&g^{\alpha_1\alpha'_n}
g^{\alpha_2\alpha'_1}g^{\alpha_3\alpha'_2}\dots
g^{\alpha_{n-1}\alpha'_{n-2}}
g^{\alpha_n\alpha'_{n-1}},\label{m1'}\\
g_{a_1\dots a_n}&=&g_{\alpha_1\alpha'_n}
g_{\alpha_2\alpha'_1}g_{\alpha_3\alpha'_2}\dots
g_{\alpha_{n-1}\alpha'_{n-2}}
g_{\alpha_n\alpha'_{n-1}}\label{m2'}.
\ee
The case $n=1$ corresponds to the Poisson tensor and its
inverse. 

\subsection{Poisson and Lie-Poisson brackets}

The Hamilton equations imply
the Poisson bracket equations
\be
i\, \partial_s F&=&-g_{\alpha\alpha'}
\Bigl(
\frac{\delta F}{\delta \psi_{\alpha}}
\frac{\delta H}{\delta\bar \psi_{\alpha'}}
-
\frac{\delta H}{\delta \psi_{\alpha}}
\frac{\delta F}{\delta\bar \psi_{\alpha'}}
\Bigr)\label{weinberg}\\
&=&
-g_{\alpha\beta'}\rho_{\beta\alpha'}
\Bigl(
\frac{\delta F}{\delta \rho_{\alpha\alpha'}}
\frac{\delta H}{\delta \rho_{\beta\beta'}}
-
\frac{\delta H}{\delta \rho_{\alpha\alpha'}}
\frac{\delta F}{\delta \rho_{\beta\beta'}}
\Bigr)
=
\rho_a\Omega^a{_{bc}}
\frac{\delta F}{\delta \rho_{b}}
\frac{\delta H}{\delta \rho_{c}}.\label{LP}
\ee
The structure constants are 
\be
\Omega^a{_{bc}}&=&
\ve_{\gamma'}{^{\alpha'}}
\ve_{\beta}{^{\alpha}}
g_{\gamma\beta'}
-
\ve_{\beta'}{^{\alpha'}}
\ve_{\gamma}{^{\alpha}}
g_{\beta\gamma'}\\
\Omega{_{abc}}&=&g_{ad}\Omega^d{_{bc}}=
-
g_{\alpha\beta'}
g_{\beta\gamma'}
g_{\gamma\alpha'}
+
g_{\alpha\gamma'}
g_{\beta\alpha'}
g_{\gamma\beta'}
\label{O_}\\
\Omega{^{abc}}&=&g^{bd}g^{ce}\Omega^a{_{de}}=
g^{\alpha\beta'}
g^{\beta\gamma'}
g^{\gamma\alpha'}
-
g^{\alpha\gamma'}
g^{\beta\alpha'}
g^{\gamma\beta'}.\label{O^}
\ee

\subsection{Lie-Nambu bracket form of linear proper time dynamics}

Denote $S=S[\rho]=S(C_2[\rho])=g^{ab}\rho_a\rho_b/2=:C_2[\rho]/2$. 
The 
triple Lie-Nambu bracket form of dynamics is
\be
i\partial_s F&=&\{F,H,S\}=\Omega{_{abc}}
\frac{\delta F}{\delta \rho_{a}}
\frac{\delta H}{\delta \rho_{b}}
\frac{\delta S}{\delta \rho_{c}}\label{LN}
\label{L-vN'}.
\ee
For $F=\rho_d$ 
Eq.~(\ref{L-vN'}) is the linear Liouville-von Neumann equation
in its proper time version provided $S$ is second-order in
$\rho_{a}$. 

\section{Nonlinear generalization}
\subsection{Casimir invariants}

Proofs of the theorems given below are analogous to those
from \cite{c2} so we do not present them. 
Denote 
$
g^{a_1\dots a_n}\rho_{a_1}\dots \rho_{a_n}=: C_n[\rho].
$
\begin{th}
\label{th1}
\be
\{C_n,C_m,\,\cdot\,\}=0.
\ee
\end{th}
$C_n$ are therefore Casimir invariants for all Lie-Nambu brackets.
\begin{th}
\label{th2}
Let $S=S(C_1,\dots C_k,\dots)$ be any
differentiable function of $C_1,\dots C_k\dots$, and $H$, $F$ arbitrary (in
general nonlinear) observables. Then
\be
\{C_n,F,S\}&=&0,\\
\partial_s C_n=-i\{C_n,H,S\}&=&0.
\ee
\end{th}

\subsection{N particles and separation of subsystems}

Let 
$g^{N}{^{ab}}=g{^{a_1b_1}}\dots g{^{a_Nb_N}}$, 
$g^{N}{_{ab}}=g{_{a_1b_1}}\dots g{_{a_Nb_N}}$.
Consider an $N$-particle density matrix 
$\rho{^N}_{a}=\rho_{a_1\dots a_N}$. In linear QM a $K$-particle subsystem
($K\leq N$) is described by observables of the form
\be
F^K&=&g^{Nab}F_{a_1\dots a_K}g_{a_{K+1}}\dots g_{a_N}
\rho_{b_1\dots b_N}=
g^{Kab}F_{a_1\dots a_K}
\rho_{b_1\dots b_K},
\ee
where 
\be
\rho_{b_1\dots b_K}&=&
g^{a_{K+1}b_{K+1}}\dots g^{a_{K}b_{N}}
g_{a_{K+1}}\dots g_{a_N}
\rho_{b_1\dots b_Kb_{K+1}\dots b_N}\nonumber\\
&=&
g^{b_{K+1}\dots b_{N}}
\rho_{b_1\dots b_Kb_{K+1}\dots b_N}
\ee
is the subsystem's reduced density matrix. Consider now two,
$M$- and $(N-M-K)$-particle, 
subsystems which do not overlap (i.e. no particle belongs to
both of them). Their reduced density matrices are 
\be
\rho{^I}_{d}&=&\rho{^I}_{d_1\dots d_M}=
\rho_{d_1\dots d_Md_{M+1}\dots d_N}
g^{d_{M+1}\dots d_N},\\
\rho{^{II}}_{e}&=&\rho{^{II}}_{e_{M+K+1}\dots e_N}=
g^{e_1\dots e_{M+K}}
\rho_{e_1\dots e_{M+K}e_{M+K+1}\dots e_N}
\ee
then 
\begin{th}
\label{th3}
\be
\{\rho{^{I}}_{d},\rho{^{II}}_{e},\,\cdot\,\}^N=0.
\ee
\end{th}
The $N$-particle triple bracket is defined in terms of the 
$N$-particle structure constants which are of the one-particle
form but now with all $g$'s replaced by $g^N$'s \cite{c2}.
Theorem~\ref{th3} implies
\begin{th}
\label{th4}
Consider two, in general nonlinear, observables 
$F^{I}[\rho]=F^{I}[\rho^{I}]$, $G^{II}[\rho]=G^{II}[\rho^{II}]$
corresponding 
to two nonoverlapping, $M$- and $(N-M-K)$-particle subsystems
of a larger $N$-particle system. Then
$
\{F^{I},G^{II},\,\cdot\,\}^N=0.
$
\end{th}
The meaning of Theorem~\ref{th4} is the following. 
Let a composite system consisting of two
noninteracting subsystems be described by a 
(possibly nonlinear) Hamiltonian function 
$
H[\rho]=H^{I}[\rho^{I}] + H^{II}[\rho^{II}]
$.
Then, for {\it any\/} $S$
$
i\partial_s F^{I}=\{F^{I}, H, S\}=\{F^{I}, H^{I}, S\}
$
and the dynamics of a subsystem is generated by the
Hamiltonian function of this subsystem. Theorem~\ref{th4} is a
general result stating that the triple-bracket scheme allows for
a consistent composition of $N$-particle dynamics from 
elementary single-particle ones. It follows that
the density matrix formalism, as opposed to the 
standard nonlinear Schr\"odinger equation pure-state framework,
does not introduce any new ``threshold phenomena" 
in transition from $N$ to $N+1$ particle systems (cf. \cite{GS}).

\section{Convexity principle and nonlinearity: A few remarks}

A density matrix is usually thought of as a kind of mixture of
fundamental (quantum) and ordinary (classical) probabilities. 
As such it is typically attributed to ensembles of many
particles as opposed to a state vector which, at least in some
interpretations, may be regarded as a property of a single system. 
This perspective suggests that a role of density matrices 
should be reduced to this of a simple mathematical tool allowing
for mixing a
classical lack of knowledge with fundamental quantum
probabilities. Mathematically this seems to imply that 
the Liouville-von Neumann equation must be linear even if pure
states evolve nonlinearly. This point of view forms an implicit 
philosophical basis of
Mielnik's formalism \cite{M} which on one hand does not exclude
nonlinear evolutions
of pure states forming the boundary of a ``figure of states",
and on the other requires the figure to be convex. 

The triple bracket formalism leads to a weaker form of the
convexity principle \cite{c3} 
which can be formulated as the following 
\begin{th}
\label{th5}
Let $\rho_0=\sum_{k=1}^\infty p_k(0)|k,0\rangle\langle k,0|$ be
a density matrix acting in a separable Hilbert space, and let 
$\rho(t)$ be a Hermitian solution of a triple bracket equation
with $H$ and $S=S(C_1,\dots,C_k,\dots)$ arbitrary.
If $\rho(0)=\rho_0$ then for any $t$ there exists a basis
${|k,t,\{p_k(0)\}\rangle}$ such that 
$\rho(t)=\sum_{k=1}^\infty p_k(0)
{|k,t,\{p_k(0)\}\rangle}
{\langle k,t,\{p_k(0)\}|}$.
\end{th}
The nonlinearity is manifested in the dependence of 
${|k,t,\{p_k(0)\}\rangle}$ on $,\{p_k(0)\}$. The
figure of states is now no longer convex but the eigenvalues of
the density matrix can be nevertheless interpreted in the standard
way. The problem arises whether one can obtain such a dynamics
in a typical counting experiment where the ensemble in question
consists of separately arriving particles. Our guess is that
this should not be the case and that the nonlinear evolution has
to correspond to more complicated physical situations. 
In a more general perspective we are inclined to depart from the
usual interpretation and regard density matrices as more 
fundamental than state vectors. Some particular cases might then
correspond to classical mixtures in analogy to the role played
in quantum mechanics by Abelian subalgebras of observables. 

\section{Appendix}

The bispinor convention we use is the following. To any Greek
index there corresponds a pair of Latin ones written
down in a lexicographic order. For example
\be
F{_{\alpha}}{^{\beta'}}{_{\gamma}}=
\left(
\begin{array}{c}
F{_{A}}{^{B'}}{_{C}}\\
F{_{A}}{^{B'}}{_{C'}}\\
F{_{A}}{^{B}}{_{C}}\\
F{_{A}}{^{B}}{_{C'}}\\
F{_{A'}}{^{B'}}{_{C}}\\
\vdots
\end{array}
\right);
\quad
{\ve}_{\alpha}{^{\beta'}}=
\left(
\begin{array}{c}
{\ve}_{A}{^{B'}}\\
{\ve}_{A}{^{B}}\\
{\ve}_{A'}{^{B'}}\\
{\ve}_{A'}{^{B}}
\end{array}
\right)=
\left(
\begin{array}{c}
0\\
\ve_{A}{^{B}}\\
\ve_{A'}{^{B'}}\\
0
\end{array}
\right).\nonumber
\ee
Any permutation preserving the lexicographic rule induces a
natural isomorphism, say, $F{_{\alpha}}{^{\beta'}}{_{\gamma}}
\to F{_{\alpha'}}{^{\beta'}}{_{\gamma}}$ where the latter 
bispinor would begin with $F{_{A'}}{^{B'}}{_{C}}$. 
In particular
\be
g_{\alpha}{^{\beta'}}
=
\left(
\begin{array}{c}
0\\
-\ve_{A}{^{B}}\\
\ve_{A'}{^{B'}}\\
0
\end{array}
\right);
\quad
g_{\alpha'}{^{\beta}}
=
\left(
\begin{array}{c}
0\\
\ve_{A'}{^{B'}}\\
-\ve_{A}{^{B}}\\
0
\end{array}
\right);
\quad
g_{\alpha}{^{\beta}}
=
\left(
\begin{array}{c}
-\ve_{A}{^{B}}\\
0\\
0\\
\ve_{A'}{^{B'}}
\end{array}
\right);
\quad
g_{\alpha'}{^{\beta'}}
=
\left(
\begin{array}{c}
\ve_{A'}{^{B'}}\\
0\\
0\\
-\ve_{A}{^{B}}
\end{array}
\right).\nonumber
\ee
The bispinor summation convention is illustrated by
$
G^\alpha H_\alpha=G^{A} H_{A}+G^{A'} H_{A'}=G^{\alpha'} H_{\alpha'}$.
\section*{Acknowledgments}
M.~C. wants to thank A.~Jadczyk for pointing out the
problem of hyperplane dependence of the triple bracket formalism
and other critical remarks. He is also grateful to G.~A.~Goldin,
K.~Jones, F.~Gaioli and E.~T.~Garcia Alvarez 
for many inspiring and fruitful discussions, and
the Conference organizers for financial support.

\end{document}